\documentclass[12pt]{article}

\begin{document}

\begin{center}
{\Large\bf{}Dominant property for the Bel-Robinson tensor and tensor S}\\
Lau Loi So\footnote{email address: s0242010@gmail.com}
\end{center}

\begin{abstract}
The Bel-Robinson tensor contains many nice mathematical properties
and its dominant energy condition is desirable for describing the
positive gravitational energy. The dominant property is a basic
requirement for the quasi-local mass, i.e., in small sphere limit.
We claim that there exists another option, a linear combination
between the Bel-Robinson tensor $B$ and tensor $S$, which
contributes the same dominant property. Moreover, using the 5
Petrov types as the verification, we found that this dominant
property justification for the Bel-Robinson tensor can be
simplified as examining $B_{0000}\geq|B_{\alpha\beta{}00}|$ and
$B_{0000}\geq|B_{0123}|$, instead of
$B_{0000}\geq|B_{\alpha\beta\lambda\sigma}|$ for all
$\alpha,\beta,\lambda,\sigma=0,1,2,3$.
\end{abstract}

\section{Introduction}
Gravitational energy cannot be localized at a point since it is
forbiddance by the equivalence principle. However, the quasi-local
method (i.e., small sphere~\cite{Horowitz-small-sphere}) can solve
out this difficulty. For describing the gravitational energy, Bel
and Robinson~\cite{Bel-1,Bel-2,Robinson-1,Robinson-2} proposed a
tensor that is positive definite and satisfy the dominant
property~\cite{Senovilla}. This dominant property is a relevant
requirement for the quasi-local mass. The Bel-Robinson tensor also
possesses other nice properties: completely symmetric, traceless
and divergence free.

The quasi-local mass has been studied for a long time. There were
many people attempted to give a definition for this subject:
Harking~\cite{Hawking}, Penrose~\cite{Penrose}, Brown and
York~\cite{BrownYork}, etc. Recently, Wang and Yau~\cite{WangYau}
proposed certain requirements and one of them is the Bel-Robinson
tensor in vacuum. Indeed, in order to obtain this positivity in
small sphere, it is believed that it should be proportional to the
Bel-Robinson tensor~\cite{Szabados}. However, would it be the only
choice?  We claim that there exists another option, a linear
combination between the Bel-Robinson tensor $B$ and tensor $S$,
which contributes the same dominant property (i.e., see
(\ref{20Dec2016})).

In principle, because of the symmetry, the Bel-Robinson tensor
contains 35 components. Using the 5 Petrov types~\cite{Petrov} as
the verification, we observe that the dominant property
justification for $B$ can be simplified as examining
$B_{0000}\geq|B_{\alpha\beta{}00}|$ and $B_{0000}\geq|B_{0123}|$,
instead of $B_{0000}\geq|B_{\alpha\beta\lambda\sigma}|$ for all
$\alpha,\beta,\lambda,\sigma=0,1,2,3$.  Because all components
contain in $B$ can be written in terms of $B_{\alpha\beta{}00}$
and $B_{0123}$.

\section{Dominant energy condition for $B+sS$}
Analogy with the theory of electrodynamics, the Bel-Robinson
tensor is defined as
\begin{eqnarray}
B_{\alpha\beta\lambda\sigma}:=C_{\alpha\xi\lambda\kappa}C_{\beta}{}^{\xi}{}_{\sigma}{}^{\kappa}
+*C_{\alpha\xi\lambda\kappa}*C_{\beta}{}^{\xi}{}_{\sigma}{}^{\kappa},
\end{eqnarray}
where $C_{\alpha\beta\mu\nu}$ is the Weyl conformal tensor and its
dual $\ast{C_{\alpha\beta\mu\nu}}=\frac{1}{2}
\epsilon_{\alpha\beta\lambda\sigma}
C^{\lambda\sigma}{}_{\mu\nu}$~\cite{Carmeli}. As the Weyl tensor
and Riemann tensor are equivalent in vacuum, the energy density
for this Bel-Robinson tensor becomes
\begin{eqnarray}
B_{\alpha\beta\lambda\sigma}t^{\alpha}t^{\beta}t^{\lambda}t^{\sigma}=E_{ab}E^{ab}+H_{ab}H^{ab},
\end{eqnarray}
which is non-negative and $t^{\alpha}$ is the timelike unit
normal. Here the Greek letters refer to spacetime and Latin stand
for space. In vacuum, the Bel-Robinson tensor and tensor
$S$~\cite{MTW} can be defined as folows
\begin{eqnarray}
B_{\alpha\beta\lambda\sigma}&:=&R_{\alpha\xi\lambda\kappa}R_{\beta}{}^{\xi}{}_{\sigma}{}^{\kappa}
+R_{\alpha\xi\sigma\kappa}R_{\beta}{}^{\xi}{}_{\lambda}{}^{\kappa}
-\frac{1}{8}g_{\alpha\beta}g_{\lambda\sigma}R^{2}_{\rho\tau\mu\nu},\\
S_{\alpha\beta\lambda\sigma}&:=&R_{\alpha\lambda\xi\kappa}R_{\beta\sigma}{}^{\xi\kappa}
+R_{\alpha\sigma\xi\kappa}R_{\beta\lambda}{}^{\xi\kappa}
+\frac{1}{4}g_{\alpha\beta}g_{\lambda\sigma}R^{2}_{\rho\tau\mu\nu},
\end{eqnarray}
where
$R^{2}_{\rho\tau\mu\nu}=R_{\rho\tau\mu\nu}R^{\rho\tau\mu\nu}$ is
the Kretschmann scalar. The symmetric property for $S$ is
$S_{\alpha\beta\lambda\sigma}=S_{(\alpha\beta)(\lambda\sigma)}=S_{\lambda\sigma\alpha\beta}$.
It is known that the Bel-Robinson tensor possesses the dominant
property
$B_{\alpha\beta\lambda\sigma}u^{\alpha}v^{\beta}w^{\lambda}z^{\sigma}\geq0$,
where $u,v,w,z$ are any future-pointing causal vectors. This
dominant energy condition is significant for defining the
quasi-local mass~\cite{WangYau}. Here we propose another option, a
linear combination between $B$ and $S$ such that it also possesses
the dominant property:
\begin{eqnarray}
(B_{\alpha\beta\lambda\sigma}+sS_{\alpha\beta\lambda\sigma})u^{\alpha}v^{\beta}w^{\lambda}z^{\sigma}\geq0,
\label{20Dec2016}
\end{eqnarray}
where $s$ is a non-zero small constant.

For the dominant property (i.e., dominant super-energy condition),
Senovilla proposed a definition (see Lemma 4.1 of
\cite{Senovilla}): ``If a tensor $T_{\mu_{1}...\mu_{s}}$ satisfies
the dominant super-energy property, then
$T_{0...0}\geq|T_{\mu_{1}...\mu_{s}}|$, $\forall$
$\mu_{1},\ldots,\mu_{s}=0,\ldots,n-1$ in any orthonormal basis
$\{\vec{e}_{\nu}\}$". For example, using the 5 Petrov types as the
examination, the Bel-Robinson tensor fulfills
$B_{0000}\geq|B_{\alpha\beta\lambda\sigma}|$ requirement.
Likewise, for $B+sS$ and we found there exists a non-zero small
$s$ such that
\begin{eqnarray}
B_{0000}+sS_{0000}\geq|B_{\alpha\beta\lambda\sigma}+sS_{\alpha\beta\lambda\sigma}|.
\end{eqnarray}
Thus, we suggest that the quasi-local mass should include this
extra candidate $B+sS$ in small sphere. Referring to Szabados's
argument~\cite{Szabados}, ``Therefore, in vacuum in the leading
$r^{5}$ order any coordinate and Lorentz-covariant quasi-local
energy-momentum expression, which is nonspacelike and future
pointing must be proportional to the Bel-Robinson `momentum'
$B_{\beta\lambda\sigma\alpha}t^{\beta}t^{\lambda}t^{\sigma}$." We
claim that $B+sS$ is not only satisfy the causal, but also the
dominant property. However, there is a disadvantage for $B+sS$
because we need to check $s$ in every physical system.
Nevertheless, the advantage for $B+sS$ gives a relaxation
opportunity since obtaining the pure Bel-Robinson tensor for a
quasi-local expression is not easy.

Here we consider the total energy-momentum complex which accurate
to zeroth order in matter and second order in empty spacetime
\begin{eqnarray}
{\cal{T}}^{\alpha}{}_{\beta}=T^{\alpha}{}_{\beta}+t^{\alpha}{}_{\beta\lambda\sigma}x^{\lambda}x^{\sigma},
\end{eqnarray}
where $T^{\alpha}{}_{\beta}$ is the stress tensor and
$t^{\alpha}{}_{\beta\lambda\sigma}$ is the gravitational
pseudotensor. Note that there are 2 free indices in
${\cal{T}}_{\alpha\beta}$.  Confining within the small sphere
region, $t^{\alpha}{}_{\beta\lambda\sigma}x^{\lambda}x^{\sigma}$
satisfies the divergence free condition~\cite{SoNesterPRD}:
$\partial_{\alpha}(t^{\alpha}{}_{\beta\lambda\sigma}x^{\lambda}x^{\sigma})=0$.
The gravitational energy-momentum in small sphere is
\begin{eqnarray}
\begin{array}{ccccc}
\int_{V}t_{\alpha\beta\lambda\sigma}x^{\lambda}x^{\sigma}d^{3}x
=\frac{4\pi}{15}(t_{\alpha\beta\lambda\sigma}\eta^{\lambda\sigma}+t_{\alpha\beta{}00})r^{5},
\end{array}
\end{eqnarray}
where we used the spherical coordinates and allow the time
component be constant for simplicity. Indeed
${\cal{T}}_{\alpha\beta}$ is symmetric in $\alpha,\beta$.
Moreover, the dominant energy condition confined in small sphere
limit is
\begin{eqnarray}
t_{\alpha\beta{}00}u^{\alpha}v^{\beta}\geq0,
\end{eqnarray}
where $t_{\alpha\beta\lambda\sigma}\eta^{\lambda\sigma}$ is an
arbitrary constant according to the symmetry. If $t$ is replaced
by $B$ and $B+sS$ respectively, we have the simplified dominant
property representation:
\begin{eqnarray}
B_{\alpha\beta{}00}u^{\alpha}v^{\beta}\geq0,\quad{}
(B_{\alpha\beta{}00}+sS_{\alpha\beta{}00})u^{\alpha}v^{\beta}\geq0.
\end{eqnarray}
The second inequality is valid for a suitable non-zero small $s$.

What is the criterion for selecting the small $s$? Here we give a
concrete example by using an isotropic Schwarzschild line element
in polar coordinates
\begin{eqnarray}
ds^{2}=-(1-2Mr^{-1})dt^{2}+(1-2Mr^{-1})^{-1}dr^{2}+r^{2}(dr^{2}+\sin^{2}\theta\,d\phi^{2}),
\end{eqnarray}
with the assumption that $M/r<<1$, both the gravitational constant
$G$ and speed of light $c$ are unity. For simplicity, using the
orthonormal basis, there are only three non-vanishing components
$(E_{11},E_{22},E_{33})=(-2,1,1)Mr^{-3}$. The value for the
quadratic scalar is $R^{2}_{\rho\tau\lambda\sigma}=48M^{2}r^{-6}$.
The non-vanishing components for $B$ and $S$ are
\begin{eqnarray}
(B_{0000},B_{0011},B_{0022},B_{0033})&=&(6,-2,4,4)M^{2}r^{-6},\nonumber\\
(B_{1111},B_{2222},B_{3333},B_{1122},B_{1133},B_{2233})&=&(6,6,6,-4,-4,2)M^{2}r^{-6},\nonumber\\
(S_{0000},S_{0011},S_{0022},S_{0033})&=&(12,-28,-16,-16)M^{2}r^{-6},\nonumber\\
(S_{0101},S_{0202},S_{0303},S_{1111},S_{2222},S_{3333})&=&(8,2,2,12,12,12)M^{2}r^{-6},\nonumber\\
(S_{1122},S_{1133},S_{2233},S_{1212},S_{1313},S_{2323})&=&(16,16,28,-2,-2,-8)M^{2}r^{-6}.
\end{eqnarray}
Obviously, the Bel-Robinson tensor fulfills the dominant energy
condition. Similarly, we find that $(B+sS)$ satisfies the dominant
property requires $s\in[-\frac{1}{14},\frac{1}{4}]$. In
particular, for the Landau-Lifschitz (LL)
pseudo-tensor~\cite{Deser}, evaluated in the Riemann normal
coordinates, satisfies the dominant property:
\begin{eqnarray}
\begin{array}{ccccc}
\partial^{2}_{\mu\nu}t^{\alpha\beta}_{LL}
=\frac{1}{9}\left(7B^{\alpha\beta}{}_{\mu\nu}+\frac{1}{2}S^{\alpha\beta}{}_{\mu\nu}\right).
\end{array}
\end{eqnarray}

\section{Conclusion}
The Bel-Robinson tensor $B$ has the dominant energy condition and
this is a requirement for describing the quasi-local mass in small
sphere. We discovered that there exists an opportunity tensor
$B+sS$ such that this combination also contributes the same
dominant property. As it is not easy for achieving a multiple of
the pure Bel-Robinson tensor in quasi-local expression, then
$B+sS$ provides a relaxation opportunity for the dominant energy
condition. Moreover, we also pointed out that the examination for
the dominant property can be simplified for the Bel-Robinson
tensor. Using the 5 Petrov types, instead of verifying
$B_{0000}\geq|B_{\alpha\beta\lambda\sigma}|$ for all
$\alpha,\beta,\lambda,\sigma=0,1,2,3$, it is enough to check
$B_{0000}\geq|B_{00\alpha\beta}|$ and $B_{0000}\geq|B_{0123}|$.


\begin{thebibliography}{3}

\bibitem{Horowitz-small-sphere}
Horowitz G T and Schmidt B G 1982 {\it Proc. Roy. Soc. Lond. A}
{\bf 381}

\bibitem{Bel-1}
Bel L 1958 {\it CR Acad. Sci. Paris} {\bf 247} 1094

\bibitem{Bel-2}
Bel L 1958 {\it CR Acad. Sci. Paris} {\bf 248} 1297

\bibitem{Robinson-1}
Robinson I 1958 unpublished Kings College Lectures

\bibitem{Robinson-2}
Robinson I 1997 {\it Class. Quantum Grav.} {\bf{}14} 4331




\bibitem{Senovilla}
Senovilla J M M 2000 {\it Class. Quantum Grav.} {\bf{}17} 2799


\bibitem{Hawking}
Hawking S W 1968 {\it J. Math. Phys} {\bf 9}



\bibitem{Penrose}
Penrose R 1982 {\it Proc. Roy. Soc. Lond. A}  {\bf 381}


\bibitem{BrownYork}
Brown J D and York J W 1993 {\it Phys. Rev. D} {\bf 47} 1407



\bibitem{WangYau}
Wang M T and Yau S T 2009 {\it Phys. Rev. lett.} {\bf 102} 021101



\bibitem{Szabados}
Szabados L B 2009 Living Rev. Rel. {\bf 12} 4



\bibitem{Petrov}
Gomez-Lobo A G P 2008 {\it Class. Quantum. Grav.} {\bf 25} 015006



\bibitem{Carmeli}
Carmeli M {\it ``Classical Fields General relativuty and Gauge
Theory"} (John Wiley $\&$ Sons 1982)







\bibitem{MTW}
Misner C W, Thorne K S and Wheeler J A 1973 {\it{}Gravitation}
(San Francisco, CA: Freeman)



\bibitem{SoNesterPRD}
So L L and Nester J M 2009 {\it Phy. Rev. D} {\bf{}79} 084028






\bibitem{Deser}
Deser S, Franklin J S and Seminaea D 1999 {\it Class. Quantum
Grav.} {\bf 16} 2815






\end{thebibliography}
\end{document}